\documentclass[conference]{IEEEtran}

\makeatletter

\def\ps@IEEEtitlepagestyle{%
  \def\@oddfoot{\mycopyrightnotice}%
  \def\@evenfoot{}%
}
\def\mycopyrightnotice{%
  {\footnotesize 979-8-3315-3559-9/25/\$31.00~\copyright~2025 IEEE\hfill}%
  \gdef\mycopyrightnotice{}
}

\usepackage{blindtext}
\usepackage{url}
\IEEEoverridecommandlockouts
\usepackage{cite}
\usepackage{amsmath,amssymb,amsfonts}
\usepackage{algorithmic}
\usepackage{graphicx}
\usepackage{textcomp}
\usepackage{xcolor}
\usepackage{listings}
\def\BibTeX{{\rm B\kern-.05em{\sc i\kern-.025em b}\kern-.08em
    T\kern-.1667em\lower.7ex\hbox{E}\kern-.125emX}}
    
\usepackage{eso-pic}
\usepackage{hyperref}

\newcommand\AtPageUpperMyright[1]{\AtPageUpperLeft{%
 \put(\LenToUnit{0.17\paperwidth},\LenToUnit{-2cm}){%
     \parbox{0.9\textwidth}{\raggedleft\fontsize{8}{11}\selectfont #1}}%
 }}%
\newcommand{\conf}[1]{%
\AddToShipoutPictureBG*{%
\AtPageUpperMyright{#1}
}
}

\begin{document}
\title{\vspace*{1cm} ``vcd2df" - Leveraging Data Science Insights for Hardware Security Research\\\thanks{This work was supported by NSF Award 171858}}

\author{\IEEEauthorblockN{Calvin Deutschbein, Jimmy Ostler, Hriday Raj}
\IEEEauthorblockA{\textit{School of Computing and Information Sciences} \\
\textit{Willamette University}\\
Salem, Oregon, United States of America \\
\{ckdeutschbein,jtostler,hhraj\}@willamette.edu}
}

\maketitle
\conf{\textit{  Proc. of International Conference on Artificial Intelligence, Computer, Data Sciences and Applications (ACDSA 2025) \\ 
7-9 August 2025, Antalya-Türkiye}}
\begin{abstract} In this work, we hope to expand the universe of security practitioners of open-source hardware by creating a bridge from hardware design languages (HDLs) to data science languages like Python and R through novel libraries that convert VCD (value change dump) files into data frames, the expected input type of the modern data science tools. We show how insights can be derived in high-level languages from register transfer level (RTL) trace data. Additionally, we show a promising future direction in hardware security research leveraging the parallelism of Spark to study transient execution CPU vulnerabilities, and provide reproducibility resources via GitHub\footnote{\url{https://github.com/vcd2df/}} and Colab\footnote{\url{github.com/vcd2df/spark/blob/main/vcd2df_spark_iflow_demo.ipynb}}.
\end{abstract}



\section{Introduction}

Modern hardware designs are exceedingly complex~\cite{schoeberl23}, on the order of 10 billion transistors for consumer CPUs (central processing unit). To combat complexity, Hardware Description Languages (HDLs) enable hardware designers to design software by writing code. To become still more accessible, we can recognize that traces of execution of a hardware design are no different than any other observations from a statistical or data scientific perspective and can be interacted with via the same methodologies.

In this work, we will introduce ``vcd2df" libraries for what we believe to be the three most common data science frameworks: the R Project for Statistical Computing, Python ``pandas", and ASF Spark (which incidentally supports Python and R frontends). For each, we will show an example of a hardware design, a hardware trace as VCD file, and an insight we can discover with language built-in functions and methods. We recognize that a simulation-only approach is insufficient for some hardware goals but still supports hardware engineering.

\begin{enumerate}
    \item \textbf{The R Project}: A ``free software environment for statistical computing and graphics."\footnote{\url{www.r-project.org/}}
    \item \textbf{Python ``pandas"}: A ``fast, powerful, flexible and easy to use open source data analysis and manipulation tool, built on top of the Python programming language."\footnote{\url{pandas.pydata.org/}}
    \item \textbf{ASF Spark}: A ``multi-language engine for executing data engineering, data science, and machine learning on single-node machines or clusters."\footnote{\url{spark.apache.org/}}
\end{enumerate}

\section{Background}

We briefly introduce the input and output types of our framework: (1) HDLs and the VCD file type, and (2) Data frames.
\subsection{HDLs and the VCD file type}

What is a ``vcd", or value change dump? ``[A]n ASCII-based format for dumpfiles generated by EDA logic simulation tools. The standard, four-value VCD format was defined along with the Verilog hardware description language by the IEEE Standard 1364-1995~\cite{Verilog} in 1996."\footnote{\url{en.wikipedia.org/wiki/Value_change_dump}}. VCD files capture traces of execution of simulated software design, for which hardware engineers often interact with the traces through domain-specific tools like waveform viewers, such as GTKWave\footnote{\url{gtkwave.sourceforge.net/}}.

VCD files are generated by software tools which simulate, emulate, or compile in a machine-executable language some hardware design, expressed in a hardware design language such as Verilog or VHDL. They commonly have three primary inputs: design, testbench, and simulator.

While many aspects of the VCD file format are unimportant, we do note that the term ``value change" refers to the fact that VCD files only log changes to the bits within some register (what are essentially a fixed number of fixed size variables in a hardware design), which is commonly expressed as binary numeric data, though VCD files allow hardware signals in bits in addition to `0' and `1', so a bit may have a value of `x' or `z'. This merited three major design decisions for these packages:

\begin{enumerate}
    \item First, we treat any register containing any non-numeric value as having a value of `-1', which is easy to encode in all data frame implementations and not a valid register value, as register values are expressed as natural numbers in binary. This deviates from the ``pandas" standard (NaN) or using nullables, but works well in all environments that support signed integers, allowing greater portability.
    \item Second, we implement our packages via file streaming, rather than loading the entire file at once into memory, as we necessarily maintain prior register values in the data frame and may read a single value change at a time. Many traces are far to large to fit into RAM in any format.
    \item Third, we noted that many VCD files contained registers of the same name in different modules, a hardware abstraction similar to the software abstraction of the same name, that never differed in value, and removed these to avoid data duplication.
\end{enumerate}

Together, we believe these design decisions balance correctness and performance.

We are not unaware of the rising popularity of the FST (Fast Signal Trace) type, but it is not currently used by our research partners and is not an open IEEE standard. We hope to extend our framework to include FST in the future as it becomes more popular in hardware security research.

\subsubsection{A Design}

The primary input is a hardware design specified in an HDL such as Verilog or VHDL. In general, we expect a design specified at the register transfer level (RTL). In the case of Verilog, these are often specific with a ``.v" suffix and specify hardware designs as a series of registers or signals joined by wires. This represents the digital logic of some hardware design, but can readily be interpreted as a program or executable.

There are numerous open-source Verilog designs made available through a wide-range of research~\cite{restuccia21} and industry~\cite{parisi24} projects.

\subsubsection{A Testbench}

An HDL description of hardware cannot be executed and therefore cannot generate a trace of execution, which is necessary to create a trace of execution and have values for logging. Therefore VCD generation also requires a testbench, a hardware level description of some operations for the design, which usually exercises some or all of the design through a series of loops and inputs. Testbench generation is a separate and active area of research~\cite{zheng24} but testbench of any kind is sufficient to develop our packages.

In general, we find that testbenches are often maintained under version control in the same HDL as the design to which they accompany, as development without testbenches is exceedingly difficult and uncommon. For all designs we explored, we were able to use existing testbenches written by project maintainers.

\subsubsection{A Simulator}

We find that the most commonly recommended tool for scholarship on hardware design is Icarus Verilog, a free and open-source Verilog compiler under the GPL license, maintained on GitHub\footnote{\url{github.com/steveicarus/iverilog}}. ``Icarus Verilog is not aimed at being a simulator in the traditional sense, but a compiler that generates code employed by back-end tools." Icarus Verilog supports the VCD file as a default output type.

We should note that the VCD format is ASCII-based format, incurring write-speed limits at one-eighth of the speed of a binary representation (when printing textual binary, one bit of information incurs eight bits of storage). In initial experiments, we have found it possible to bypass the VCD representation and stream directly into a data science framework, but this removes the benefit of encapsulating hardware design complexities entirely. We hope to explore streaming data during hardware simulation in future work.

\subsection{Data Frames}

What is a ``df", or data frame? ``[O]ne of the most common data structures used in modern data analytics because they are a flexible and intuitive way of storing and working with data."\footnote{\url{databricks.com/glossary/what-are-dataframes}}. Today, data frames are the state-of-the-art for data analysis but lacked a way to be applied to hardware engineering and security.

Rather than being the result of an IEEE standard like VCD, the notion of a data frame emerged naturally over time as statistical, scientific computing, and data analytics libraries became more mature in scripting languages and cloud computing applications. We are aware of no precise definition or history around data frames, but think of them as a base feature of the R Project, which was formally launched in 1993, and incorporated into Python by ``pandas" (for ``panel data") in 2008 and then by Spark 1.6 in 2015. Statisticians developing R noted the usefulness of storing data with variables in columns and cases, or observations, in rows. In 2025, ``pandas" (and its dependency NumPy) are the two most popular libraries in the most popular programming language (Python) and Spark is one of the most popular platforms for distributed computing.

We formally regard a data frame as a ``tabular data structure common to many data processing libraries"\footnote{\url{en.wikipedia.org/wiki/Dataframe}} and think of them as fulfilling many of the goals of spreadsheet applications with the benefits of the data structure being accessible as an object within a scripting environment. It is useful as a technology to scale data analysis beyond graphical user interfaces and facilitate automation.

\subsubsection{The R Project}

The R Project uses the term ``data frame" to refer to lists of the ``data.frame" class, and (following arrays, matrices, and lists) is the most basic data structure in R which is not common to other programming languages. Essentially, they are implemented as lists of vectors, which may be lists, array, martices or other data frames.

\subsubsection{Python ``pandas"}

The ``pandas" library regards the DataFrame as its primary data structure. The DataFrame is implemented as a dictionary, or map, like structure over ``Series" objects, its vector type. The vectors are implemented as NumPy arrays, which are in turn implemented as arrays in the C programming language.

The usefulness of this implementation for using data frames for hardware design is that that the similarly designed devices with 32 or 64 bit integers perhaps implicitly act as self-hosted hardware accelerators for study of hardware designs. The NumPy and Pyarrow backing of ``pandas" both offer a true-to-hardware encoding many common register sizes.

\subsubsection{ASF Spark}

Spark and its underlying insight, resilient distributed datasets (RDDs)~\cite{zaharia12}, was introduced in 2014 to address limitations of the earlier paradigm of MapReduce~\cite{dean08} and was an important innovation in distributed computing. 

A core motivation for this work was to develop frameworks that scale to any hardware designs, including CISC (Complex Instruction Set Computer) CPUs, SoCs (System-on-a-Chip), and other hardware that would be impractical to simulate and analyze on a single device. For example, automated testing for transient execution CPU vulnerabilities (such as Spectre~\cite{kocher18} or Meltdown~\cite{lipp18}) is a long-standing goal of hardware security research~\cite{deutschbein21} but requires extensive computation. We begin this process in our case studies.

\section{Implementation}

A variety of existing tools interact with VCD files, yet to our knowledge no currently supported tools read VCD files into data frames of any kind. However, we were able to adapt work from the Mythra\footnote{\url{github.com/hwcicd/myrtha}} package for containerized hardware CI/CD~\cite{deutschbein25}, which used a custom Python script to parse VCD files for specification mining, and instead target data frame output. We implemented the VCD-to-data-frame parsing natively in both Python, with a ``pandas" dependency, and in R with no dependencies, so that the package may be easily maintained and distributed.

We elected for native implementation versus the use of low-level language such as C/C++ or Rust, as we experience a performance bottleneck on HDL simulation rather than on our parsing for all designs. However, if performance in these libraries becomes a bottleneck on design, we expect the most likely improvements to come from streaming value changes with no intermediate write-to-disk or ASCII encoding, which took on the order of hours~\cite{deutschbein21} (versus minutes) on our example designs.

Both packages provide the 'vcd2df' function, which loads a IEEE 1364-1995/2001 
VCD (.vcd) file, specified as an parameter as a string containing exactly 
a file path, and returns either a pandas DataFrame or an R Project list of class data.frame 
containing values over time.  
A VCD file captures the register values at discrete timepoints from a 
simulated trace of execution of a hardware design in Verilog or VHDL.
The returned data frame contains a row for each register, by name, and a
column for each time point, specified VCD-style using octothorpe-prefixed 
multiples of the timescale.

\subsubsection{The R Project}

As part of research efforts, we have published with CRAN (The Comprehensive R Archive Network) a ``vcd2df" package\footnote{\url{cran.r-project.org/web/packages/vcd2df}} which implements a single function of the same name. We implemented the package natively in R in 86 lines-of-code, of which 33 were comments and 3 were whitespace only.

\subsubsection{Python ``pandas"}

As part of research efforts, we have published to PyPI (The Python Package Index) a ``vcd2df" package\footnote{\url{pypi.org/project/vcd2df/1.0/}} which implements a single function of the same name. We implemented the package natively in Python in 36 lines-of-code, of which 3 were comments and 3 were whitespace only.  We may use this same package in ASF Spark via the PySpark interface, or use this package to preprocess VCD files into Parquet files for direct usage in Spark.

\section{Case Studies}

We tested over three distinct hardware designs, using one technology per design. We used the demonstration designs for the Mythra package which include two RISC-V~\cite{asanovic14} Reduced Instruction Set Computers (RISC) provided open-source by Yosys and an access control module~\cite{restuccia21} provided open-source by the Kastner Research Group (KRG) at UCSD:

\begin{enumerate}
    \item NERV - Naive Educational RISC-V Processor\footnote{\url{github.com/YosysHQ/nerv}}
    \item PicoRV32 - A Size-Optimized RISC-V CPU\footnote{\url{github.com/YosysHQ/picorv32}}
    \item AKER-Access-Control\footnote{\url{github.com/KastnerRG/AKER-Access-Control}}
\end{enumerate}

Each design was written in Verilog, simulated with Icarus Verilog, and open-sourced on GitHub. We used Icarus Verilog version 11 rather than the most recent release (12) to avoid a versioning issue with the latest Verilog standard and some non-standard implementation features of NERV. While these files can be generated from open-source designs with open-source tools, we also maintain a repository with the VCD files\footnote{\url{github.com/vcd2df/vcd_ex}} for accessibility and reproducibility.

We timed translations via the `time' command and the average and standard deviation from 10 translations, after writing minimal scripts that wrapped the ``vcd2df" library import and function call. In R, we wrote:

\begin{lstlisting}[language=R]
library(vcd2df)

args <- commandArgs()
f_name <- args[length(args)]
df <- vcd2df(paste0(f_name, ".vcd"))
saveRDS(df, paste0(f_name, ".rds"))
\end{lstlisting}

In Python, the "pandas" I/O API supports many filetypes, including both text-based and binary files, and with the addition of "pyreadr" module\footnote{\url{https://pypi.org/project/pyreadr/}} can also write to the R Project RDS filetype. In Python, we wrote:

\begin{lstlisting}[language=Python]
from vcd2df import vcd2df
import sys
    
f_name = sys.argv[1]
df = vcd2df(f_name+".vcd")
df.to_pickle(f_name+".pkl")
\end{lstlisting}

We should note we tested both the ``pyarrow" and ``fastparquet" engines for Parquet files. Parquet and Feather are ASF file types and well suited to Spark (and usable in R packages with appropriate R packages), and ``pickle" is a Python file type, essentially the Python equivalent of the R Project RDS file type.

\subsection{NERV}

NERV contains 549 registers in 1267 lines of SystemVerilog. Its accompanying testbench is 155 SystemVerilog LoC (lines of code) and runs for 19 cycles. In R, translation took an average of .194 seconds with a standard deviation of .005 seconds. In Python, translation to ``pickle" took an average of .303 seconds with a standard deviation of .006 seconds. We report the file sizes and savings ratios in Tab.~\ref{tab:nerv}. We remark briefly on the space savings, for example, the VCD file was 31.0 kilobytes in size but the resultant RDS file was only 2.51 kilobytes in size, a space savings of 93\%.

\begin{table}[h]
\caption{NERV data frame sizes}
\centering
\begin{tabular}{lrrrrr}
File Type & File Ext. & Engine & Bytes & Save\% & ms \\
VCD & .vcd & iverilog & 31049 & - & - \\
RDS & .rds & R & 2512 & 93.0 & 194 \\
pickle & .pkl & pandas & 35926 & -8.72 & 312 \\
Parquet & .parquet & PyArrow & 8629 & 76.0 & 352 \\
Parquet & .parquet & Fast & 6646 & 81.5 & 309 \\
Feather & .ftr & PyArrow & 9786 & 72.8 & 318 \\
RDS & .rds & pyreadr & 30057 & 16.3 & 318 \\
\end{tabular}
\label{tab:nerv}
\end{table}

To show the usefulness of data frames for studying NERV traces, we note that the space savings can be unexpectedly high, so we may wish to assess the coverage of our testbench. We can do so with a few simple lines of R:

\begin{lstlisting}[language=R]
library(vcd2df)
nerv <- vcd2df("nerv.vcd")
count <- data.frame(rowSums(nerv > 0))
colnames(count) <- "counts"
rownames(subset(count, counts > 0))
 \end{lstlisting}

We can quickly see that only 20 registers (of 549) are initialized and set to non-zero values in this brief test, and can additionally see their names, including noteworthy registers such as `trap', `clock', and `reset', common classes of names for registers defining the control state of CPU designs like NERV. Notably, `pc', the program counter, is not included yet usually is the most important register for denoting progression through some trace of execution as it traces successive instructions read from memory.

\subsection{PicoRV32}

PicoRV32 contains 232 registers in 3049 lines of Verilog code. Its accompanying testbench is 86 lines of Verilog and runs for 2201 cycles. In R, translation took an average of 1.618 seconds with a standard deviation of .024 seconds. In Python, translation to "pickle" took an average of .601 seconds with a standard deviation of .047 seconds. We report the file sizes and savings ratios in Tab.~\ref{tab:pico}. We remark briefly on the space cost, for example, the VCD file was 269 kilobytes in size but the resultant "pickle" file was 3963 kilobytes in size, a space cost of 14.6$\times$.

\begin{table}[h]
\centering
\caption{PicoRV32 data frame sizes}
\begin{tabular}{lrrrrr}
File Type & File Ext. & Engine & Bytes & Size$\times$ & ms \\
VCD & .vcd & iverilog        & 269184   & -     & - \\
RDS & .rds & R               & 107012   & .398  & 1618 \\
pickle & .pkl & pandas       & 3917766  & 14.6  & 601 \\
Parquet & .parquet & PyArrow & 1704664  & 6.33  & 706 \\
Parquet & .parquet & Fast    & 1191528  & 4.42  & 1533 \\
Feather & .ftr & PyArrow     & 1717122  & 6.37  & 681 \\
RDS & .rds & pyreadr         & 3962508  & 14.7  & 1150 \\
\end{tabular}
\label{tab:pico}
\end{table}

To show the usefulness of data frames for studying PicoRV32 traces, we note that the ``32" in name stands for the word size, and we can examine registers to see if each contains what we expect to be memory addresses - values smaller than $2^{32}$ which are also multiples of 32. We term this set $M$.
$$
M = \{ m \in \mathbb N : (32 \mid m) \land (m < 2^{32}) \}
$$
It is trivial to implement the membership test in Python, both for individual values and for collections of values, as either functions or lambda expressions containing the following code blocks:

\begin{lstlisting}[language=Python]
not (m % 32) and (m < (1 << 32)) # single
all([f(m) for m in ms]) # collection
 \end{lstlisting}

From there, we construct a series of registers for which this predicate holds at all time points.

\begin{lstlisting}[language=Python]
df = pd.read_pickle("pico.pkl")
df.apply(f_all, axis=1)
 \end{lstlisting}

We can quickly see that only 56 registers (of 221) are 32 bit values which are multiples of 32 at all time points, including some memory registers such as `mem\_la\_firstword'. However, not all `mem' prefixed registers, like `mem\_wstrb', are in this series, suggesting we should further study our assumptions.

\subsection{AKER}

AKER contains 432 registers in two files totaling 2002 Verilog LoC. Its accompanying testbench is 527 lines of SystemVerilog and runs for 1055 cycles. In R, translation took an average of .423 seconds with a standard deviation of .010 seconds. In Python, translation to "pickle" took an average of .440 seconds with a standard deviation of .007 seconds. We report the file sizes and savings ratios in Tab.~\ref{tab:aker}.

\begin{table}[h]
\centering
\caption{AKER data frame sizes}
\begin{tabular}{lrrrrr}
File Type & File Ext. & Engine & Bytes & Size$\times$ & ms \\
VCD & .vcd & iverilog        & 53007    & -     & - \\
RDS & .rds & R               & 56969    & 1.07  & 423 \\
pickle & .pkl & pandas       & 2977387  & 56.1  & 440 \\
Parquet & .parquet & PyArrow & 888735   & 16.8  & 518 \\
Parquet & .parquet & Fast    & 684618   & 12.9  & 949 \\
Feather & .ftr & PyArrow     & 952394   & 18.0  & 489 \\
RDS & .rds & pyreadr         & 2997098  & 56.5  & 859 \\
\end{tabular}
\label{tab:aker}
\end{table}

For AKER, to exercise Spark, we use 229 VCD files generated by the Isadora tool for automated information flow property generation for hardware designs~\cite{deutschbein21}, which open-sourced its evaluation data via GitHub\footnote{\url{github.com/cd-public/Isadora/tree/master/model/single/vcds}}. Each VCD file tracks potential sensitive information from one for the 229 registers and denotes an information flow into another register by setting a register of the same name prefixed with ``shadow\_" to a non-zero value. We slightly reworked the ``vcd2df" function to read strings as ``strd2df", rather than files, and read the VCD files into a Spark Dataframe as text data then convert to a ``pandas" DataFrame.

\begin{lstlisting}[language=Python]
vcds = "gs://vcds-aker/vcds/*.vcd"
df = spark.read.text(vcds, wholetext=True)
rdd = df.rdd.map(lambda x : str2df(x[0]))
 \end{lstlisting}

From there, we could straightforwardly inspect ``shadow\_" registers, registers which denote information flow tracking information, to find the first non-zero instance (when sensitive information may reach a register).

\begin{lstlisting}[language=Python]
df = df[df.index.str.contains("shadow")]
df = df[df.any(axis=1)]
df = df.idxmax(axis=1)
 \end{lstlisting}

We find the data frame contains what we believe to be previously undescribed information flow paths. To the best of our knowledge, this has been achieved by no other existing techniques for trace analysis. We have made a few extended Spark scripts available via GitHub\footnote{\url{github.com/vcd2df/spark}}, including a Colab\footnote{\url{github.com/vcd2df/spark/blob/main/vcd2df_spark_iflow_demo.ipynb}} demo and a more Spark-like implementation with ``flatMap".

We note that we are underutilizing Spark at this scale, and executing on a single cluster only as a proof-of-concept. Approaching a fully distributed approach to Spark would require significant additional investment in VCD generations as well as processing, but we hope to approach in future research.

\section{Related Work}

\paragraph*{Python VCD Interfaces}

There are existing interfaces from Python to VCD files, namely PyVCD\footnote{\url{https://pypi.org/project/vcdvcd/}} and vcdvcd\footnote{\url{https://pypi.org/project/pyvcd/}}. Versus our tool, PyVCD only writes VCD files, and vcdvcd only supports ASCII formatted output, which scales poorly to moderately sized designs. We elected to convert directly to binary encoding for scalability of write and query operations.

\paragraph*{Data Scientific Approaches to Hardware}
Early computational methods to assess the correctness and security of hardware designs often directly examined trace data for known patterns, such as one-hot encoding~\cite{hangal05,mandouh12}. Other researchers applied data mining~\cite{chang10,hertz13,li10} or temporal logic~\cite{danese2017team,danese2015automaticdate,danese2015automatic,deutschbein18} techniques. Specific the to security context, researchers have manually identified properties~\cite{hicks15,irvine11,brown17} and used software tools incorporating clustering and classification for RISC~\cite{ZhangASPLOS2017} and CISC~\cite{deutschbein20} designs. Recent work has discovered subsets of hyperproperties~\cite{rawat20,deutschbein21}, which are defined as relationships over multiple traces. We use Spark to evaluate multiple traces of a single design (PicoRV32) to discover grammatically similar hyperproperties to these works.

\paragraph*{Data Scientific Approaches to Correctness}
Ammons et al. introduced specification mining~\cite{ammons02}, a form of data mining for computing systems by considering their traces of execution, and did so in a software context. Specification mining launched a rich research direction across static and dynamic analysis~\cite{weimer05}; imperfect traces ~\cite{yang06}; and complex races~\cite{gabel2008javert,reger13,gabel2008symbolic}. Perhaps the most widely known miner, Daikon~\cite{ernst07} approached specification mining as invariant detection. An inspiration for this work was studying the Daikon source code and seeing custom implementations of $k$-means and hierarchical clustering used within its heuristics for determining program correctness. We regard the Python and R implementations of these algorithms as the state-of-the-art and wished to apply them directly to hardware traces.

\paragraph*{Python Testbenches}

The use of languages common to data analytics for hardware evaluation is not novel, with perhaps the most common example being specifying testbenchs in Python~\cite{sharma19} which allows study of hardware design while abstracting register transfer level notions of sequential and combinatorial logic and non-numerical register values.

Pyverilog~\cite{takamaeda15} was an earlier effort to bring Verilog tooling, such as code generation and code flow analysis, into a higher-level language.

We believe both of these contributions take advantage of the accessibility of Python as a scripting language, but do not leverage its unique capabilities for data analysis.

\section{Conclusion}

We have proposed and demonstrated ``vcd2df", a proof-of-concept package for studying hardware designs with existing data science techniques. We have distributed our package through the relevant package managers for Python and the R Project, and have shown simple examples of insights from data frames containing trace data. Research is ongoing to study transient execution CPU vulnerabilities, and initial results suggest improved expressibility versus the state-of-the-art specification miners.

\section*{Acknowledgements}

We thank the NSF for funding. We thank the CRAN network for help developing ``vcd2df" for the R Project. We thank Andres Meza for the AKER testbench. We thank Intel Corporation, the Kastner Research Group, and the HWSec@UNC research group for helpful discussions. We thank Kendall Leonard and Hannah Pahama for suggesting library support for a VCD dataframe in their summer REU of 2024.

\bibliographystyle{unsrt}
\bibliography{refs}

\begin{thebibliography}{10}

\bibitem{schoeberl23}
Martin Schoeberl.
\newblock Open-source hardware design.

\bibitem{Verilog}
Ieee standard verilog hardware description language.
\newblock {\em IEEE Std 1364-2001}, pages 1--792, 2001.

\bibitem{restuccia21}
Francesco Restuccia, Andres Meza, and Ryan Kastner.
\newblock Aker: A design and verification framework for safe and secure soc access control.
\newblock In {\em 2021 IEEE/ACM International Conference On Computer Aided Design (ICCAD)}, pages 1--9, 2021.

\bibitem{parisi24}
Emanuele Parisi, Alberto Musa, Maicol Ciani, Francesco Barchi, Davide Rossi, Andrea Bartolini, and Andrea Acquaviva.
\newblock Assessing the performance of opentitan as cryptographic accelerator in secure open-hardware system-on-chips, 2024.

\bibitem{zheng24}
Ziyue Zheng, Xiangchen Meng, and Yangdi Lyu.
\newblock Ape-fv: Concolic testing for rtl functional verification using adaptive path exploration.
\newblock In {\em 2024 IEEE 42nd International Conference on Computer Design (ICCD)}, pages 373--380, 2024.

\bibitem{zaharia12}
Matei Zaharia, Mosharaf Chowdhury, Tathagata Das, Ankur Dave, Justin Ma, Murphy McCauley, Michael~J. Franklin, Scott Shenker, and Ion Stoica.
\newblock Resilient distributed datasets: a fault-tolerant abstraction for in-memory cluster computing.
\newblock In {\em Proceedings of the 9th USENIX Conference on Networked Systems Design and Implementation}, NSDI'12, page~2, USA, 2012. USENIX Association.

\bibitem{dean08}
Jeffrey Dean and Sanjay Ghemawat.
\newblock Mapreduce: simplified data processing on large clusters.
\newblock {\em Commun. ACM}, 51(1):107–113, January 2008.

\bibitem{kocher18}
Paul Kocher, Jann Horn, Anders Fogh, , Daniel Genkin, Daniel Gruss, Werner Haas, Mike Hamburg, Moritz Lipp, Stefan Mangard, Thomas Prescher, Michael Schwarz, and Yuval Yarom.
\newblock Spectre attacks: Exploiting speculative execution.
\newblock In {\em 40th IEEE Symposium on Security and Privacy (S\&P'19)}, 2019.

\bibitem{lipp18}
Moritz Lipp, Michael Schwarz, Daniel Gruss, Thomas Prescher, Werner Haas, Anders Fogh, Jann Horn, Stefan Mangard, Paul Kocher, Daniel Genkin, Yuval Yarom, and Mike Hamburg.
\newblock Meltdown: Reading kernel memory from user space.
\newblock In {\em 27th {USENIX} Security Symposium ({USENIX} Security 18)}, 2018.

\bibitem{deutschbein21}
Calvin Deutschbein, Andres Meza, Francesco Restuccia, Ryan Kastner, and Cynthia Sturton.
\newblock Isadora: Automated information flow property generation for hardware designs.
\newblock In {\em Proceedings of the 5th Workshop on Attacks and Solutions in Hardware Security}, ASHES '21, page 5–15, New York, NY, USA, 2021. Association for Computing Machinery.

\bibitem{deutschbein25}
Calvin Deutschbein and Aristotle Stassinopoulos.
\newblock "test, build, deploy" -- a ci/cd framework for open-source hardware designs, 2025.

\bibitem{asanovic14}
K.~Asanovi{\'c} and D.~A. Patterson.
\newblock Instruction sets should be free: the case for {RISC-V}.
\newblock Technical Report UCB/EECS-2014-146, Department of Electrical Engineering and Computer Science, University of California, Berkeley, Berkeley, CA, USA, August 2014.

\bibitem{hangal05}
Sudheendra Hangal, Sridhar Narayanan, Naveen Chandra, and Sandeep Chakravorty.
\newblock {IODINE}: A tool to automatically infer dynamic invariants for hardware designs.
\newblock In {\em Proceedings of 42nd Design Automation Conference (DAC)}. IEEE, 2005.

\bibitem{mandouh12}
E.~{El Mandouh} and A.~G. {Wassal}.
\newblock Automatic generation of hardware design properties from simulation traces.
\newblock In {\em International Symposium on Circuits and Systems (ISCAS)}, pages 2317--2320. IEEE, 2012.

\bibitem{chang10}
Po-Hsien Chang and Li~C Wang.
\newblock Automatic assertion extraction via sequential data mining of simulation traces.
\newblock In {\em Proceedings of the 15th Asia and South Pacific Design Automation Conference (ASP-DAC)}, pages 607--612. IEEE, 2010.

\bibitem{hertz13}
Samuel Hertz, David Sheridan, and Shobha Vasudevan.
\newblock Mining hardware assertions with guidance from static analysis.
\newblock {\em IEEE Transactions on Computer-Aided Design of Integrated Circuits and Systems}, 32(6):952--965, 2013.

\bibitem{li10}
Wenchao Li, Alessandro Forin, and Sanjit~A Seshia.
\newblock Scalable specification mining for verification and diagnosis.
\newblock In {\em Proceedings of the 47th design automation conference}, pages 755--760, 2010.

\bibitem{danese2017team}
A.~{Danese}, N.~D. {Riva}, and G.~{Pravadelli}.
\newblock {A-TEAM}: Automatic template-based assertion miner.
\newblock In {\em Proceedings of the 54th Design Automation Conference (DAC)}, pages 1--6. ACM/EDAC/IEEE, June 2017.

\bibitem{danese2015automaticdate}
A.~{Danese}, T.~{Ghasempouri}, and G.~{Pravadelli}.
\newblock Automatic extraction of assertions from execution traces of behavioural models.
\newblock In {\em Design, Automation Test in Europe Conference Exhibition (DATE)}, pages 67--72, March 2015.

\bibitem{danese2015automatic}
A.~{Danese}, G.~{Pravadelli}, and I.~{Zandonà}.
\newblock Automatic generation of power state machines through dynamic mining of temporal assertions.
\newblock In {\em Design, Automation Test in Europe Conference Exhibition (DATE)}, pages 606--611, March 2016.

\bibitem{deutschbein18}
Calvin Deutschbein and Cynthia Sturton.
\newblock Mining security critical linear temporal logic specifications for processors.
\newblock In {\em 2018 19th International Workshop on Microprocessor and SOC Test and Verification (MTV)}, pages 18--23, 2018.

\bibitem{hicks15}
Matthew Hicks, Cynthia Sturton, Samuel~T. King, and Jonathan~M. Smith.
\newblock {SPECS}: A lightweight runtime mechanism for protecting software from security-critical processor bugs.
\newblock In {\em Proceedings of the Twentieth International Conference on Architectural Support for Programming Languages and Operating Systems (ASPLOS)}, pages 517--529. ACM, 2015.

\bibitem{irvine11}
M.~Bilzor, T.~Huffmire, C.~Irvine, and T.~Levin.
\newblock Security checkers: Detecting processor malicious inclusions at runtime.
\newblock In {\em International Symposium on Hardware-Oriented Security and Trust (HOST)}, pages 34--39. IEEE, June 2011.

\bibitem{brown17}
Michael Brown.
\newblock Cross-validation processor specifications.
\newblock Master's thesis, University of North Carolina at Chapel Hill, 2017.

\bibitem{ZhangASPLOS2017}
Rui Zhang, Natalie Stanley, Christopher Griggs, Andrew Chi, and Cynthia Sturton.
\newblock Identifying security critical properties for the dynamic verification of a processor.
\newblock In {\em Proceedings of the Twenty-Second International Conference on Architectural Support for Programming Languages and Operating Systems (ASPLOS)}, pages 541--554. ACM, 2017.

\bibitem{deutschbein20}
Calvin Deutschbein and Cynthia Sturton.
\newblock Evaluating security specification mining for a cisc architecture.
\newblock In {\em 2020 IEEE International Symposium on Hardware Oriented Security and Trust (HOST)}, pages 164--175, 2020.

\bibitem{rawat20}
Mayank Rawat, Sujit~Kumar Muduli, and Pramod Subramanyan.
\newblock Mining hyperproperties from behavioral traces.
\newblock In {\em 2020 IFIP/IEEE 28th International Conference on Very Large Scale Integration (VLSI-SOC)}, pages 88--93, 2020.

\bibitem{ammons02}
Glenn Ammons, Rastislav Bod\'{\i}k, and James~R. Larus.
\newblock Mining specifications.
\newblock In {\em Proceedings of the 29th ACM SIGPLAN-SIGACT Symposium on Principles of Programming Languages}, POPL '02, page 4–16, New York, NY, USA, 2002. Association for Computing Machinery.

\bibitem{weimer05}
Westley Weimer and George~C. Necula.
\newblock Mining temporal specifications for error detection.
\newblock In {\em Proceedings of the 11th International Conference on Tools and Algorithms for the Construction and Analysis of Systems (TACAS)}, pages 461--476. Springer-Verlag, 2005.

\bibitem{yang06}
Jinlin Yang, David Evans, Deepali Bhardwaj, Thirumalesh Bhat, and Manuvir Das.
\newblock Perracotta: Mining temporal {API} rules from imperfect traces.
\newblock In {\em Proceedings of the 28th International Conference on Software Engineering (ICSE)}, pages 282--291. ACM, 2006.

\bibitem{gabel2008javert}
Mark Gabel and Zhendong Su.
\newblock Javert: Fully automatic mining of general temporal properties from dynamic traces.
\newblock In {\em Proceedings of the 16th International Symposium on Foundations of Software Engineering (FSE)}, pages 339--349. ACM, 2008.

\bibitem{reger13}
G.~{Reger}, H.~{Barringer}, and D.~{Rydeheard}.
\newblock A pattern-based approach to parametric specification mining.
\newblock In {\em 28th International Conference on Automated Software Engineering (ASE)}, pages 658--663. IEEE/ACM, 2013.

\bibitem{gabel2008symbolic}
Mark Gabel and Zhendong Su.
\newblock Symbolic mining of temporal specifications.
\newblock In {\em Proceedings of the 30th International Conference on Software Engineering (ICSE)}, pages 51--60. ACM, 2008.

\bibitem{ernst07}
Michael~D. Ernst, Jeff~H. Perkins, Philip~J. Guo, Stephen McCamant, Carlos Pacheco, Matthew~S. Tschantz, and Chen Xiao.
\newblock The {Daikon} system for dynamic detection of likely invariants.
\newblock {\em Science of Computer Programming}, 69(1-3):35--45, December 2007.

\bibitem{sharma19}
Varun Sharma, Naif Tarafdar, and Paul Chow.
\newblock Sonar: Writing testbenches through python.
\newblock In {\em 2019 IEEE 27th Annual International Symposium on Field-Programmable Custom Computing Machines (FCCM)}, pages 311--311, 2019.

\bibitem{takamaeda15}
Shinya Takamaeda-Yamazaki.
\newblock Pyverilog: A python-based hardware design processing toolkit for verilog hdl.
\newblock In {\em Applied Reconfigurable Computing}, volume 9040 of {\em Lecture Notes in Computer Science}, pages 451--460. Springer International Publishing, Apr 2015.

\end{thebibliography}

\vspace{12pt}

\end{document}